\documentclass[10pt,conference,fleqn]{IEEEtran}
\IEEEoverridecommandlockouts
\usepackage{cite}
\usepackage{amsmath,amssymb,amsfonts}
\usepackage{algorithmic}
\usepackage{graphicx}
\usepackage{caption,subcaption}
\usepackage{adjustbox}
\usepackage[table,xcdraw]{xcolor}
\usepackage{textcomp}
\usepackage{multirow}
\usepackage{multicol}
\usepackage{booktabs}
\usepackage{bigstrut}
\usepackage{verbatim}
\usepackage{array}
\usepackage{rotating}
\usepackage{xcolor}
\usepackage{dblfloatfix}
\usepackage[a4paper, total={184mm,239mm}]{geometry}
\AtBeginDocument{%
  \providecommand\BibTeX{{%
    \normalfont B\kern-0.5em{\scshape i\kern-0.25em b}\kern-0.8em\TeX}}}
\renewcommand{\baselinestretch}{0.99}
\begin{document}

\title{Delay Balancing with Clock-Follow-Data: Optimizing Area Delay Trade-offs for Robust Rapid Single Flux Quantum Circuits  \\
}

\author{\IEEEauthorblockN{Robert S. Aviles, Phalgun G K, Peter A. Beerel}
\IEEEauthorblockA{\textit{Department of Electrical and Computer Engineering, University of Southern California}, Los Angeles, USA \\\{rsaviles, kariyapp, pabeerel\}@usc.edu}}


\maketitle

\begin{abstract}
This paper proposes an algorithm for synthesis of clock-follow-data designs that provides robustness against timing violations for RSFQ circuits while maintaining high performance and minimizing area costs. Since superconducting logic gates must be clocked, managing data flow is a challenging problem that often requires the insertion of many path balancing D Flips (DFFs) to properly sequence data, leading to a substantial increase in area. 
To address this challenge, we present an algorithm to insert DFFs into clock-follow-data RSFQ circuits that partially balances the delays within the circuit to achieve a target throughput while minimizing area. 
Our algorithm can account for expected timing variations and, 
by adjusting the bias of the clock network and clock frequency, 
we can mitigate unexpected timing violations post-fabrication. 
Quantifying the benefits of our approach with a benchmark suite with nominal delays, our designs offer an average 1.48x improvement in area delay product (ADP) over high frequency full path balancing (FPB) designs and a 2.07x improvement in ADP over the state of the 
art robust circuits provided by state-of-the-art (SOTA) multi-phase clocking solutions.  
%

\end{abstract}

\section{Introduction}

Superconductive single-flux quantum devices (SFQ) \cite{isvlsi2} have emerged as a promising technology that can provide low power exascale supercomputing \cite{Sergey2016} with a theoretical potential of three orders of magnitude power reduction compared to SOTA semiconductor devices \cite{thesis5}.  SFQ has the potential for many applications in high speed digital computation, spanning from a recently the taped-out 32GHz microprocessor \cite{32GMP} to accelerating spiking neural networks \cite{Karamuftuoglu_2024}.  Furthermore, due to operating at 4.2K, SFQ can offer high performance in-fridge computing for superconducting quantum computers, critically reducing both the bandwidth and number of wires in and out of the cryostat.  SFQ applications to quantum computing has included qubit control \cite{DigiQ}, rapid power-efficient error correction\cite{SFQ_ECC}, and moving classical portions of hybrid algorithms in-fridge \cite{SFQVQE,SFQQAOA} reducing up to 97\% of the bandwidth back to room temperature logic. Given the importance of hybrid classical-quantum algorithms \cite{2022hybrid,mcclean2016theory} SFQ may play a critical role in the scalability of quantum computers. 


\begin{figure}[t]
\includegraphics[width=0.8\columnwidth]{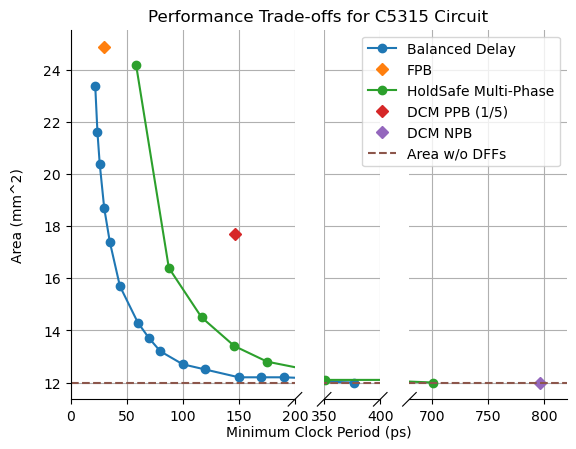}
\centering
\caption{Performance vs area curve for our work compared to existing methods on C5315 benchmark circuit}
\label{fig:Sweep}
\end{figure}
However, the achievable complexity of superconducting devices is currently limited by numerous challenges. In particular, SFQ device physics requires that all logic gates must be clocked \cite{isvlsi4}, producing deep gate-level pipelines that must be carefully synchronized to ensure proper operation. Their gate-level pipelined nature motivated many approaches to require fanin paths to be length-matched, forcing the insertion of many path balancing D Flip-Flops (DFFs) which substantially increases area and power consumption, where in some instances inserted DFFs can be up to 4.5x the original gate count \cite{Ghasem}.  

The clocked nature of each gate also results in small logic delays between clocked elements making SFQ circuits particularly vulnerable to hold violations, where as in CMOS, a single hold violation can make the chip inoperable regardless of clock frequency. This problem is exacerbated by the significant process variations currently present in the developing techniques of superconducting fabrication \cite{Rami8,Rami9}.

While inspiration can be drawn from the principles of previous CMOS methods \cite{CarverMead,WavePipeliningT}, the unique device features of SFQ logic prompt novel solutions designed to address these challenges and utilize the clocked nature of gates in optimization. As such, several novel clocking 
methodologies have recently been proposed \cite{Ghasem,isvlsi3,AvilesMultiPhase,Mingye}.  However, despite  their tailored optimization techniques, due to the complexity of SFQ data synchronization, these techniques all still require significant sacrifices in one or more of the key areas of robustness, circuit area, or performance.

Thus motivated, we propose a new clocking algorithm that optimizes area for a given target throughput while still providing robustness to timing variations. Our algorithm can exceed the high performance of FPB designs \cite{PBMap} with smaller area and achieve similar area savings of advanced approaches that use more than one clock~\cite{Ghasem,ISVLSI} at much higher performance. Figure~\ref{fig:Sweep} illustrates our improvements in area and delay for a range of target clock periods for one benchmark circuit.

\begin{figure*}[t] 
\begin{subfigure}[t]{0.33\textwidth}
\includegraphics[width=0.7\linewidth]
{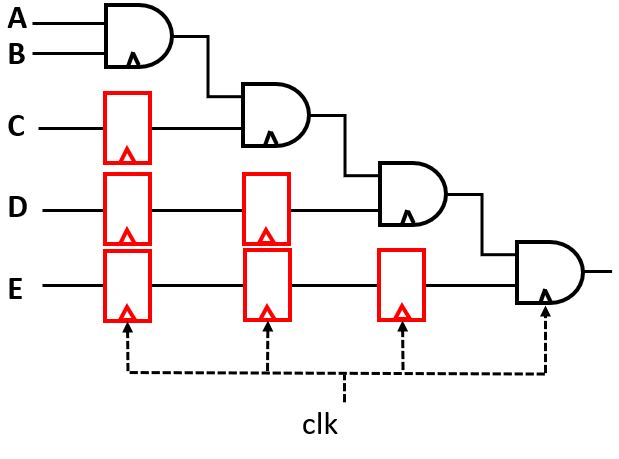}   \caption{Full path balancing: Numerous DFFs are inserted to levelize the circuit}
  \label{fig:AND_FPB}
\end{subfigure}
~
\begin{subfigure}[t]{0.33\textwidth}
\includegraphics[width=0.7\linewidth]{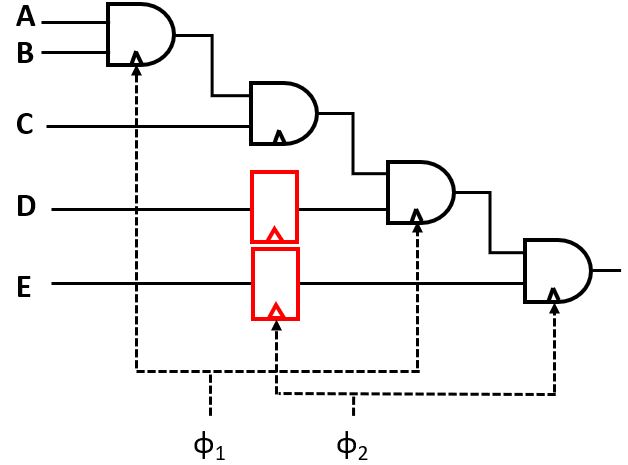}  
  \caption{Multi-phase clocking with two phases 
  $\phi_1$ \& $\phi_2$ enabling partial path balancing.}
  \label{fig:AND_MPC}
\end{subfigure}
~
\begin{subfigure}[t]{0.33\textwidth}
\includegraphics[width=0.7\linewidth]{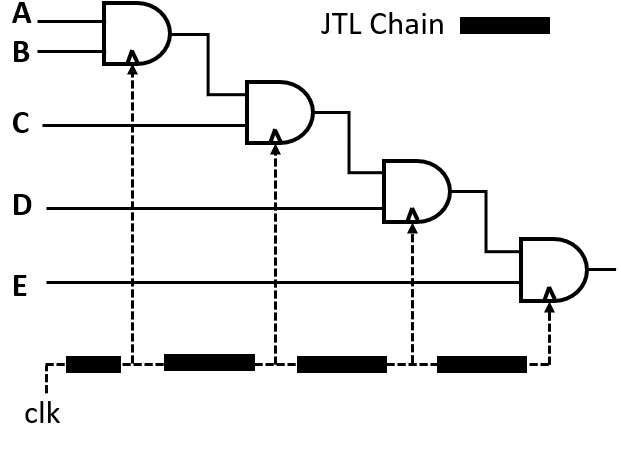}  
  \caption{Clock-follow-data: Delayed clock signal arrives after the data at each gate.}
  \label{fig:AND_DPB}
  \end{subfigure}
  \label{clocking_types}
  \caption{Circuit implementation for 5-bit AND under different clocking schemes}
\end{figure*}
\section{Delay Balancing Scheme Overview}

Our clocking scheme operates under the principles of clock-follow data \cite{isvlsi3}, but in contrast to other works \cite{CFDMargins}, we remove the need to levelize the network with inserted DFFs and do not rely on Josephson transmission lines (JTL) to balance delays along datapaths. Instead, we use added DFFs to {\em partially} balance the delays of the circuit and schedule the clock arrival times to each gate to reach our target throughput with minimal area. Our optimization algorithm can account for a range of expected delay values and supports mitigating timing violations post-fabrication. Our methodology is evaluated on common benchmark circuits validated using a simulation-based testbench to ensure functional correctness and that all timing checks are met. Our contributions can be summarized as follows:
\begin{itemize}
    \item We propose a synthesis algorithm to optimize area for a given clock frequency using clock-follow-data. When targeting high frequencies our method achieves an average 1.48x improvement in area delay product (ADP) over FPB designs without sacrificing robustness. 
    \item  Compared to multi-phase clocking \cite{AvilesMultiPhase},
    the state of the art for robust low-area circuit design, our work can achieve similar area and robustness, while operating at a much higher throughput, leading to a 2.07x improvement in ADP.
    \item  Lastly, we demonstrate how our algorithm supports uncertainty in arrival times, allowing for process-variation-aware optimization. Moreover, by guaranteeing our clock network has a distinct bias network, we argue that by adjusting this bias and slowing down the clock frequency remaining timing violations (including hold violations) can be mitigated post-fabrication.
\end{itemize}  


\section{Background}

\subsection{SFQ Devices}
Single flux quantum (SFQ) logic devices rely on superconducting material that show zero electrical resistance when cooled below their characteristic temperature.  When cooled to 4.2 Kelvin, the data pulses travel at 1/3 the speed of light, using passive superconductive microstrip lines
\cite{isvlsi3}.  Despite this high frequency and the cooling overhead, the devices in SFQ circuits still consume significantly less power compared to CMOS devices \cite{thesis9}.  

SFQ logic computes using magnetic flux and the associated current pulses to represent logic values. These pulses have voltage amplitudes $\approx$1 mV and last on the order of picoseconds.  A logic $1$ is defined as a pulse arriving during the given clock period, while a $0$ is the absence of any pulse through this period, illustrating the important role of the clock in SFQ computation and the subsequent difficulty associated with managing timing of each logic gate. 

\subsection{SFQ Clocking Methods}

The style of clock network has significant impact on reliability, throughput, and area of the circuit. To ensure proper functionality, data flowing through all reconvergent paths must be properly synchronized, with each clocking style providing different means of synchronization.

The most common method is full path balancing (FPB) Fig.~\ref{fig:AND_FPB}, which levelizes the logic elements by inserting DFFs such that each path from a primary input to a given logic gate contains the same number of clocked elements, leading to a high DFF overhead. In this method the data advances one logic stage each clock cycle and deep gate-level pipelining enables high frequency operation. 
However, timing uncertainties on the clock network
can increase the potential for irreparable hold violations,
requiring the insertion of hold-time buffers in the form of Josephson Transmission lines (JTLs) \cite{XiHold}. 

Dual clocking methods (DCM) \cite{Ghasem,Mingye} have been proposed where two clocks, one slow and one fast, have been used to remove path balancing DFFs at the cost of reducing throughput.  This approach trades performance for area but does not mitigate the potential dangers of hold violations as it still requires a high speed clock. More recently, multi-phase clocking (MPC) was developed \cite{ISVLSI} to reduce the DFF overhead associated with FPB.  By correct assignment of a driving clock phase to each gate, data synchronization can be maintained with partially imbalanced paths as seen in Fig.~\ref{fig:AND_MPC}, however the associated throughput drops with $N$, the number of clock phases used. 
Subsequent works have shown that a multi-phase system can be made robust to hold time failure by allowing for adjustments to the clock period to solve hold time issues post-fabrication \cite{AvilesMultiPhase}.  

Clock-follow-data \cite{isvlsi3} circuits use JTLs on the clock line to delay the arrival time of the clock to gates connected further down the line as illustrated in Fig.~\ref{fig:AND_DPB}. The amount of clock delay between connected gates is set to be greater than the propagation delay between the two gates, ensuring data arrives before the associated clock signal and resulting in a clock pulse propagating with data through the entire circuit. 


For ease of design, typically designers insert DFFs to fully balance logic levels and then set the clock delay for each stage using JTLs on the datapath, further increasing area overhead.  These JTLs ensure that despite the differing propagation delays and setup times within a logic stage, all data arrives at the next stage exactly the setup time before the clock, allowing a theoretical minimum clock period of setup plus hold \cite{isvlsi3}.  
However, under the presence of process variation and timing uncertainty, techniques typically require that significant timing slack be added, lowering peak performance, as recently discussed in \cite{CFDMargins}.  Furthermore, the lack of an automated EDA synthesis algorithm has limited the deployment of this clocking scheme to small circuits.  

To address these issues, as well as the absence of an SFQ clocking scheme that offers robustness while providing competitive area delay trade-offs, we propose a novel delay balancing algorithm for clock-follow-data circuits, as described below.  

\section{Proposed Delay-Balancing Method}
\subsection{Data Sequencing}

As a type of clock-follow-data, our clocking scheme uses JTLs to delay the arrival time of the clock signal to each gate. Subsequently a single clock pulse pushes data through the entire circuit, while allowing multiple waves of computation to propagate through the circuit simultaneously. To achieve this style of wave pipelining, the delays along datapaths must be properly managed to achieve a target clock frequency.  

Similarly, many works have shown that delay balancing along reconvergent paths is crucial for achieving high performance of CMOS based wave pipelining \cite{WavePipeliningT,WaePipeOG}. While CMOS delay balancing techniques rely on transistor sizing and delay element insertion to balance combinational paths, SFQ device characteristics can enable a higher degree of control over delays as data will not propagate through a gate without a clock signal.  Accordingly, our algorithm leverages general CMOS principles, while exploiting the clocked nature of superconducting logic elements.

The synchronous nature of each gate also introduces timing requirements. Namely, each gate must have all data inputs for a given wave arrive at least setup ($\sigma_i$) before the clock signal arrives at gate $i$ and the data tokens in the next wave have to arrive at least hold ($\eta_i$) after the clock signal. 

For two connected gates to satisfy setup constraints the delay between clock arrival times must be at least the clock-to-Q ($\beta$) delay of the driving gate plus the setup of the receiving gate. Thus, where the delayed clock arrival time at gate \textit{i} is referenced as 
$C_i$ the general setup requirements for a direct connection between gates \textit{i} \& \textit{j}, without splitters on their pathway can be described as: 
\begin{equation} \label{eq:GenSetup}
    C_j-C_i \geq \beta_i + \sigma_j 
\end{equation}
A hold time violation is when data arrives on the inputs of a gate sufficiently close to the previous clock signal, such that the desired outputs of the previous computation may be corrupted by the arriving data. Accordingly, in clock-follow-data schemes hold time violations are limited to instances of differing waves of computation colliding at a given node.  The requirement then is that the clock delay difference between connected gates must be sufficiently large to ensure a new wave will arrive at a gate after the hold safe exit of the previous wave. 
Since each new wave is inserted according to the clock period $\tau$ the hold constraint can be defined:
\begin{equation} \label{eq:GenHold}
    C_j + \eta_j \leq (C_i+\tau)+ \beta_i
\end{equation}

\begin{figure}[tb] 
\begin{subfigure}[t]{0.24\textwidth}
\centering
\includegraphics[width=1.0\linewidth]
{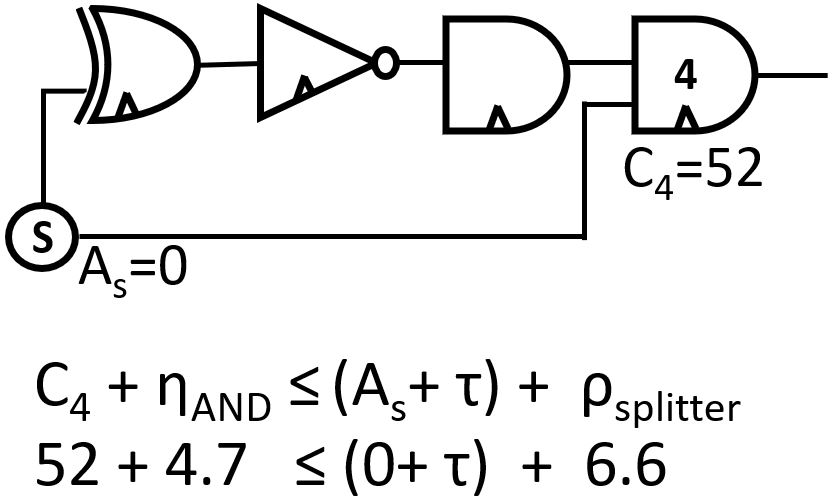}   
\caption{Delay imbalances limits clock frequency which is constrained by the splitter to AND connection. (Min $\tau$=50.1ps Table \ref{cells})}
  \label{imbal}
\end{subfigure}
\begin{subfigure}[t]{0.24\textwidth}
\centering
\includegraphics[width=1.0\linewidth]{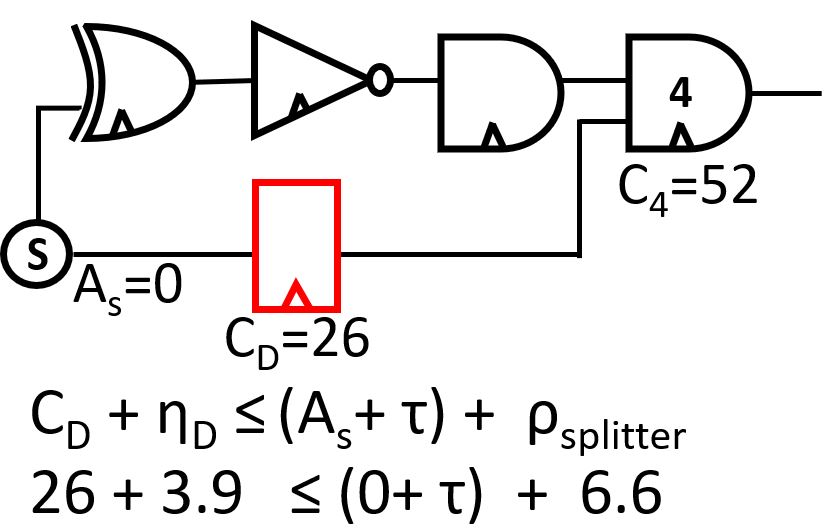}  
  \caption{Inserting a DFF improves delay balance between connections. (Min $\tau$=23.3ps)}
  \label{blanced}
\end{subfigure}
\caption{Example of reconvergent fanout showing the impact of DFF Insertion}
\label{need4dff}
\end{figure}

\begin{figure}[tb] 
\begin{subfigure}[t]{0.24\textwidth}
\centering
\includegraphics[width=1.0\linewidth]{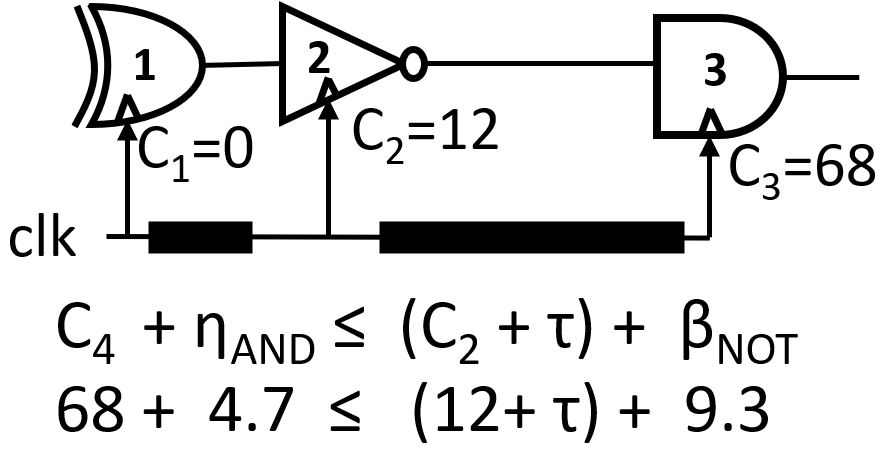}
\caption{Showing how imbalance in clock delays between gates limits throughput. (Min $\tau$ = 51.4ps Table \ref{cells})}
  \label{fig:asap}
\end{subfigure}
\begin{subfigure}[t]{0.24\textwidth}
\centering
\includegraphics[width=1.0\linewidth]{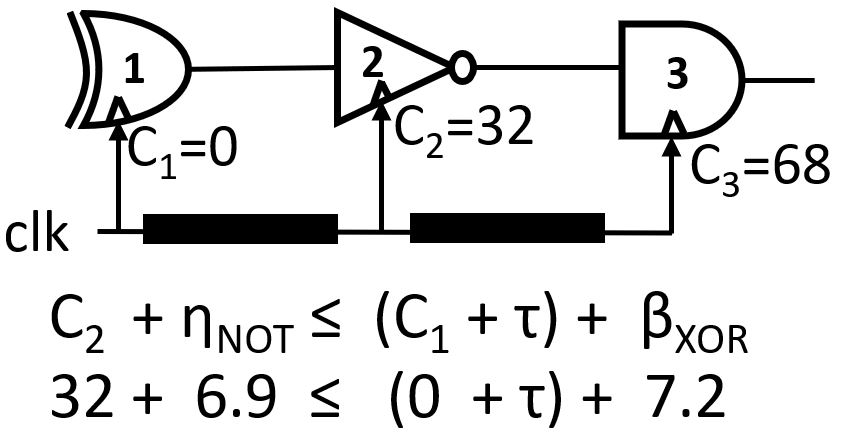} 
  \caption{Optimally delaying the clock of the NOT gate balances delay between connections.  (Min $\tau$ = 31.7ps.) }
  \label{schedule}
\end{subfigure}
\caption{Example of benefits of clock delay scheduling.}
\label{clock_schedule}
\end{figure}

This equation indicates the importance of delay balancing in achieving a minimized clock period because when $C_j$ is much larger than $C_i$ the clock period must be increased to avoid data collisions. Unfortunately, Eq. ~\ref{eq:GenSetup} constrains the minimal difference between clock delays, so for large unbalanced reconvergent paths the difference in clock delays may become too large to achieve the target clock period as shown in Fig.~ \ref{need4dff}.  However, inserting DFF(s) along this pathway can effectively increase the delay along a path, avoiding hold time collisions at our desired frequency. Additionally, as shown in Fig.~ \ref{clock_schedule}, optimizing the clock arrival times can have a critical impact on our clock period and accordingly the number of DFFs required to achieve our target frequency. 

\subsection{Post-Fabrication Timing Adjustments}

Eqs. ~\ref{eq:GenSetup} \& ~\ref{eq:GenHold} also describe the robustness of our system to unexpected timing variations. In our proposed scheme, the clock delay ($C_i$) is realized via a chain of JTLs to achieve the desired delay for all clock arrival times.  
By dedicating a separate bias line for the JTLs, designers can modify the clock delays without adjusting the parameters of the logic gates. By changing the bias current, the delay value of all JTLs can be increased allowing $C_j-C_i$ to become sufficiently large to resolve unexpected setup violations. It should be noted that this global increase in clock delay values will affect hold timing, possibly requiring an increase in clock period.  However, examining Eq.~ \ref{eq:GenHold} we can see that for any configuration of clock delays or timing variations a sufficiently large $\tau$ exists to satisfy hold constraints. Accordingly, regardless of timing variations, by decreasing the clock frequency post-fabrication we can always ensure that waves are sufficiently spaced apart to resolve any hold violations as well.  

\section{Algorithmic Implementation}

Minimizing DFFs insertion to satisfying timing encounters two key algorithmic challenges.  First, a discrete number of JTLs must be inserted, complicating optimization. To address this we add slack of 1 JTL to our constraints so that delays can be rounded up to the next highest realizable delay.  
Second, and more critically, inserted DFFs may add varying amounts of delay based on the clock delay assignment of the inserted DFF.  This requires optimization to consider not only DFF insertion but also the range of delays that each potentially inserted DFF may provide, which we define as the extra delay of a DFF.  

\begin{figure}[tbph] 
\begin{subfigure}[htbp]{0.24\textwidth}
\centering
\includegraphics[width=1.0\linewidth]
{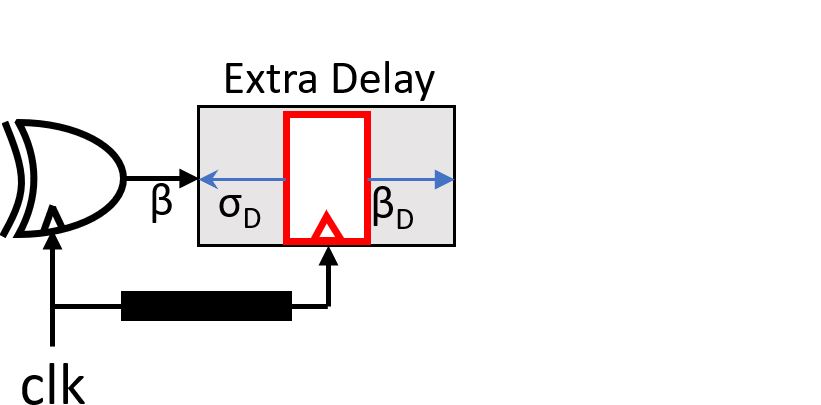}   \caption{Min extra delay of a DFF.  Delay constrained by the setup of the DFF 
 $\sigma_D$. }
  \label{MinXD}
\end{subfigure}
\begin{subfigure}[htbp]{0.24\textwidth}
\centering
\includegraphics[width=1.0\linewidth]{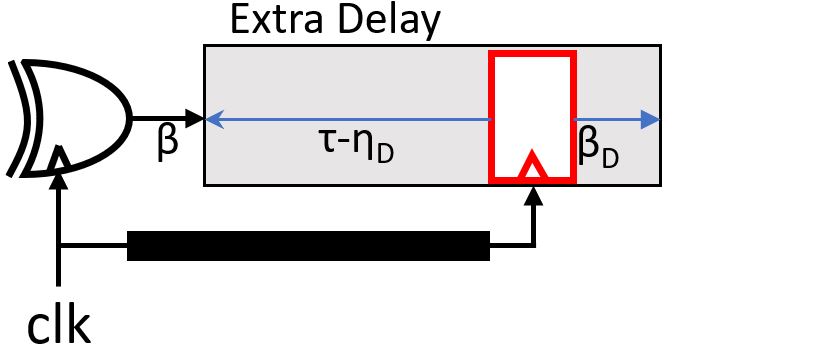}  
  \caption{Max extra delay of a DFF. Delay constrained by the period $\tau$ and hold time of the DFF $\eta_D$ }
  \label{MaxXD}
\end{subfigure}
\caption{Range of extra delays that can be added by a DFF depending on the clock delay}
\label{DFFXD}
\end{figure}

In particular, we model this extra delay as being inserted after $\beta_i$ of the driving gate and ending after the clock-to-Q time of the DFF ($\beta_{D}$).  At a minimum this extra delay must include the setup of the DFF and the clock-to-Q of the DFF Fig.~\ref{MinXD}. The maximum delay a single DFF can provide can be derived by considering the maximum clock delay of an inserted DFF using Eq. ~\ref{eq:GenHold}.   
\begin{equation} \label{eq:DFFMax}
     C_D \leq (\tau-\eta_{D}) + (C_i + \beta_i) 
\end{equation}
Since our extra delay is considered to start after $(C_i + \beta_i)$ and to end 
$\beta_{D}$ after the DFF is clocked, the maximum delay is $\tau-\eta_{D} + \beta_{D}$ (Fig.~\ref{MaxXD}).  Then the extra delay added by a chain of DFFs between gates \textit{i} \& \textit{j} can be bounded as:
\begin{equation} \label{eq:xDGen}
    F_{ij}(\sigma_{D} + \beta_{D}) \leq D_{ij} \leq F_{ij}(\tau-\eta_{D}+ \beta_{D})
\end{equation}
Here, $F_{ij}$ is the number of DFFs added and $D_{ij}$ is the extra delay provided by the inserted DFFs. Note that Eq. ~\ref{eq:xDGen} remains unchanged if the DFFs are driven by a splitter or a gate. 

In general, when a connection is driven by a splitter we can treat the splitter as if it were a gate.  In this abstraction, the propagation delay of the splitter ($\rho_i$) acts as the clock-to-Q ($\beta$) and the data arrival time on the input of the splitter ($A_i$) as the value of the clock delay. This data arrival time can be directly calculated by adding propagation delays with the clock-to-Q of the last clocked element. This level of abstraction captures wave pipelining of our splitter tree, which is possible because multiple waves of data can simultaneous travel in different levels of the tree. 
Accordingly, our algorithm includes the ability to add DFFs within a splitter tree to optimize the data arrival time at any splitter. 


With our general timing requirements in place we can model our circuit as a directed acyclic graph (DAG) where each gate or splitter is an internal node in the graph and circuit inputs (outputs) are input $PI$ (output $PO$) nodes and a directed edge ( $(i,j) \in E$ ) is created between each pair of nodes that is connected in the circuit.   The set of all gates is $G$ and the set of all splitter is $S$. Each node in $G$ is to be assigned a clock delay $C_i$ relative to a common source clock considered as delay $0$.  Each node in $S$ is assigned a data arrival time $A_i$ and has a propagation delay $\rho_i$. We partition $E$ into 4 sets, representing connection types because the variables change depending on whether the source/sink is a splitter or gate.  Specifically, splitters have arrival times and propagation delays, while gate require clock delay and have setup, hold, and clock-to-Q values.  The resulting subsets of $E$ are $GG$ for gate to gate connections, $GS$ for gate to splitter, $SG$ for splitter to gate, and $SS$ for splitter to splitter.  Regardless of subset, all connected edges are to be assigned a weight $F_{ij}$, corresponding to the number of DFFs inserted between the given nodes. Then for a given target clock period $\tau$ the optimization problem becomes: 
\begin{equation} \label{eq:objective}
    \text{Minimize:} \sum_{(i,j) \in E}{F_{ij}} \text{\hspace*{3mm} subject to:}
\end{equation}
\begin{equation}  \label{DMin}
F_{ij}(\sigma_{D} + \beta_{D} + 0.5\omega) \leq D_{ij} \hfill \forall (i,j) \in E
\end{equation}
\begin{equation} \label{DMax}
    D_{ij} \leq F_{ij}(\tau-\eta_{D}+ \beta_{D}+0.5\omega) \hfill \forall (i,j) \in E
\end{equation}
\begin{equation}\label{gtgsetup}
    C_j-C_i\geq \beta_i+\sigma_j + \omega + D_{ij} \hfill \forall (i,j) \in GG
\end{equation}
\begin{equation}\label{gtghold}
    C_j + \eta_j\leq (C_i+\tau)+ \beta_i - \omega + D_{ij} \hfill \forall (i,j) \in GG
\end{equation}
\begin{equation}\label{stgsetup}
    C_j-A_i\geq \rho_i+\sigma_j + \omega + D_{ij}  \hfill \forall (i,j) \in SG 
\end{equation}
\begin{equation}\label{stghold}
    C_j + \eta_j\leq (A_i+\tau)+ \rho_i - \omega + D_{ij} \hfill \forall (i,j) \in SG 
\end{equation}
\begin{equation}\label{gts}
    A_j = C_i+ \beta_i + D_{ij}  
    \hfill \forall (i,j) \in GS
\end{equation}
\begin{equation}\label{sts}
    A_j = A_i+ \rho_i + D_{ij} \hfill \forall (i,j) \in SS
\end{equation}
\begin{equation}\label{PI}
    C_i = 0 \hfill  \forall i \in PI 
\end{equation}
\begin{equation}\label{PO}
    C_m = C_n \hfill  \forall m,n \in P0 
\end{equation}
Here, Eqs. ~\ref{DMin} \& ~\ref{DMax} enforce the bounds on extra delay from inserted DFFs; Eqs. ~\ref{gtgsetup} \& ~\ref{gtghold} enforce the setup and hold constraints for a gate to gate connection; Eqs. ~\ref{stgsetup} \& ~\ref{stghold} enforce the setup and hold constraints for a gate driven by a splitter; and Eqs. ~\ref{gts} \& ~\ref{sts} fix the arrival times of data to a splitter driven by a gate and another splitter respectively.  Eqs. ~\ref{PI} \& ~\ref{PO} enforce all primary inputs/outputs to receive clock signals at the same time for ease of interface with other circuits.
We show the symbol definitions in Table \ref{symbols}.

\section{Algorithmic Enhancements}
\subsection{Incremental DFF insertion}\label{Incremental}
Since DFFs must be inserted in discrete numbers the optimization problem can be optimally solved as an mixed integer linear program (MILP) where each $F_{ij}$ is forced to be an integer. However, this can have prohibitive algorithmic complexity and, hence, we approximate the solution with an iterative linear program (LP) guaranteeing polynomial complexity of the whole algorithm. In this approximation, DFFs are incrementally inserted into the DAG as part of set $G$ until the desired clock period is achieved. 
Since relaxing the integer constraints on $F_{ij}$ requires rounding, the achievable timing is modified when whole number DFFs are inserted. Accordingly, the achievable timing can be checked by setting all $F_{ij}=0$ so only DFFs currently in the circuit can be considered and then adding $\tau$ to the objective function (Eq. ~\ref{eq:objective}). This will quickly return the minimal achievable clock period for the given circuit. Until the desired period is achieved, we iteratively run our DFF insertion algorithm, with the output DAG of one iteration being the input to the next.

\begin{table}[b]
  \centering
  \caption{Symbol Definitions}
  \begin{adjustbox}{max width=1.0\columnwidth}
    \begin{tabular}{|c|c|c|c|c|}
    \hline
    \textbf{Input Type} &
      \textbf{Symbol} &
      \textbf{Name } &
      \textbf{Symbol} &
      \textbf{Name }
      \bigstrut\\
    \hline
    \multirow{4}[8]{*}{\textbf{Constant}} &
      $\sigma$ &
      Setup &
      $\eta$ &
      Hold
      \bigstrut\\
\cline{2-5}     &
      $\rho$ &
      Propagation  &
      $\beta$ &
      Clock-to-Q
      \bigstrut\\
\cline{2-5}     &
      $\tau$ &
      Target Period &
      $\omega$ &
      JTL Delay
      \bigstrut\\
\cline{2-5}     &
      $\Delta^+$ &
      Increased Delay &
      $\Delta^-$ &
      Decreased Delay
      \bigstrut\\
    \hline
    \hline
    \multirow{2}[4]{*}{\textbf{Optimized Variable}} &
      $F_{ij}$ &
      Number of DFFs &
      $D_{ij}$ &
      Delay from DFFs
      \bigstrut\\
\cline{2-5}     &
      $C$ &
      Clock Delay  &
      $A$ &
      Splitter Arrival Time
      \bigstrut\\
    \hline
    \end{tabular}%
  \label{symbols}%
  \end{adjustbox}
\end{table}%

\begin{table}[t]
  \centering
  \caption{Cell Library}
    \begin{tabular}{|l|l|l|l|}
    \hline
    \textbf{Gate} &
      \textbf{Setup} &
      \textbf{Hold} &
      \textbf{Clock-to-Q}
      \bigstrut\\
    \hline
    OR &
      6.0ps &
      3.0ps &
      5.9ps
      \bigstrut\\
    \hline
    AND &
      4.5ps &
      4.7ps &
      5.7ps
      \bigstrut\\
    \hline
    XOR &
      6.9ps &
      5.5ps &
      7.2ps
      \bigstrut\\
    \hline
    NOT &
      4.4ps &
      6.9ps &
      9.3ps
      \bigstrut\\
    \hline
    DFF &
      4.4ps &
      3.9ps &
      7.9ps
      \bigstrut\\
    \hline
    \textbf{Gate} &
      \textbf{Setup} &
      \textbf{Hold} &
      \textbf{Propagation Delay}
      \bigstrut\\
    \hline
    Splitter &
      N/A &
      N/A &
      6.6 ps
      \bigstrut\\
    \hline
    JTL &
      N/A &
      N/A &
      2 ps
      \bigstrut\\
    \hline
    \end{tabular}
  \label{cells}%
\end{table}%
%
%
%

To handle our LP returning partial values for $F_{ij}$ we utilize a rounding threshold $r$. If the decimal portion of $F_{ij}$ is greater than $r$, $F_{ij}$ is rounded up to a whole integer otherwise $F_{ij}$ is rounded down. The value of $r$ is set to 1 initially and reduced by 0.2 for the first 5 iterations.  This was experimentally found to give better performance than a constant value of $r$ because it limits the DFF insertion during the initial stages when the delays will have large imbalances, effectively inserting only the most impactful DFFs until allowing progressively more aggressive insertion as needed to meet our desired clock period.  Our method is guaranteed to always return a functional circuit as throughout every iteration our DAG can construct an optimized netlist, albeit intermediate solutions are at higher clock periods than desired.  For our experiments we set a limit of 15 iterations and our method reached our target clock periods without exceeding this limit.  

Due to requiring increased slack from rounding clock delays up to the next multiple of JTL delay, our clock period will be slightly higher than the wave pipelined theoretical minimum of setup plus hold.  Our minimum becomes increased by two times the delay of a JTL.
Additionally, due to approximating DFF insertion we find that meeting this exact minimum period often incurs high DFF costs.  Hence, in our experiments we opt for increasing the target clock period to a few picoseconds above this minimum for best performance, still achieving higher clock frequency than FPB.   

\subsection{Optimizing for Expected Variations}
Designers can also optimize for a range of timing variations based on estimations of process variation or desired bias margins.  To account for these variations the expected maximum increase in delay $\Delta^+$ for each path should be considered for setup conditions, and the expected maximum decrease in delay $\Delta^-$ should be accounted for in hold.  For example, Eq.~\ref{gtgsetup} becomes Eq.~\ref{gtgnew} shown below.
\begin{equation}\label{gtgnew}
    C_j-C_i\geq \beta_i+ \Delta^+ +\sigma_j + \omega + D_{ij} \hspace*{0.4in} \hfill \forall (i,j) \in GG
\end{equation}
Similarly, the arrival time at splitters needs to consider the latest ($A_i^+$) and earliest ($A_i^-$) arrival times, where setup constraints use the latest arrival and hold constraints use the earliest arrival.  The splitter arrival times in Eq.~\ref{gts} \& \ref{sts} then become: 
\begin{equation}
    A_j^+ = C_i+ \beta_i + \Delta^+ + D_{ij} \hspace*{0.9in} \hfill \forall (i,j) \in GS
\end{equation}
\begin{equation}
    A_j^- = C_i+ \beta_i - \Delta^- + D_{ij} \hspace*{0.9in}\hfill \forall (i,j) \in GS
\end{equation}
\begin{equation}
    A_j^+ = A_i^+ + \rho_i + \Delta^+ + D_{ij} \hspace*{0.9in}\hfill \forall (i,j) \in SS
\end{equation}
\begin{equation}
    A_j^- = A_i^- + \rho_i - \Delta^- + D_{ij} \hspace*{0.9in}\hfill \forall (i,j) \in SS
\end{equation}
With these process variations accounted for our iterative DFF insertion proceeds in the same manner.  In Section \ref{margin} we compare the performance of this variation optimization to the average performance that could be obtained by post-fab timing adjustment of circuits not optimized for variation.  

\section{Experimental Results}
Our synthesis results obtained through implementation in qPALACE \cite{qPALACE} and verified with both a static timing analysis for the expected delays and verilog simulation of 1,000 waves of computation of random data inserted at the specified clock frequency.  This verilog simulation emulates the timing delays while performing setup/hold checks and verifying that the circuit outputs are functionally equivalent to the original CMOS version of the benchmarks selected.  We used common benchmark circuits \cite{EPFL,ISCAS,ShoupAlgo} of medium size as consistent with other SFQ works \cite{Ghasem,CompoundGates,AvilesMultiPhase}. In our experiments we use a cell library defined in qPALACE, with the timing values shown in Table \ref{cells}.

\subsection{Baseline Comparisons}
 As can be seen from Table \ref{VsFPB}, our design can achieve much higher clock frequencies than FPB at similar or lower area.  This leads to an average ADP Gain of 1.48x.
 
Furthermore, can substantially reduce area when our target frequency is lowered.  For comparisons to low-area designs we compare to a 3-phase hold-safe implementation of multi-phase clocking from \cite{AvilesMultiPhase}, as 3-Phases is the most competitive implementation. Given Multi-phase has been shown, in several works \cite{ISVLSI,AvilesMultiPhase,Rassul}, to significantly outperform other low-area state of the art methods, such as Dual-Clocking methods \cite{Ghasem,Mingye}, this table only directly compares to multi-phase for low-area designs in the interest of space. A target clock period of 40ps was selected because it led to similar area as the multi-phase designs with only a 0.7\% average increase in area for our designs despite running at over twice the speed.  As a result, we achieve an average of 2.07x improvement in ADP.

To illustrate the range of design options enabled by our work we present in Fig.~\ref{fig:Sweep} the area performance trade offs for the C5315 circuit, where data-points for Multi-Phase and DCM were calculated using reported DFF counts and throughput drops relative to FPB from \cite{Ghasem,AvilesMultiPhase}. 
The results show that we have smaller area at all frequencies and achieve higher maximum performance than any other method (of 22ps). Moreover, we achieve the minimum area (with no extra DFFs) at 376ps, which is 1.87x faster than the minimum area of Multi-phase and 2.1x faster than the No Path Balancing version of DCM.   
\begin{table}[htbp]
  \centering
  \caption{High Frequency Comparision to FPB}
  \begin{adjustbox}{max width=0.9\columnwidth}
    \begin{tabular}{|l|r|r|r|r|r|r|r|}
    \hline
    \multicolumn{1}{|c|}{\multirow{3}[6]{*}{\textbf{Circuit}}} &
      \multicolumn{7}{c|}{\textbf{Method}}
      \bigstrut\\
\cline{2-8}     &
      \multicolumn{3}{c|}{\textbf{Full Path Balancing}} &
      \multicolumn{4}{c|}{\textbf{Delay Balancing}}
      \bigstrut\\
\cline{2-8}     &
      \multicolumn{1}{c|}{\textbf{T}} &
      \multicolumn{1}{c|}{\textbf{DFFs }} &
      \multicolumn{1}{c|}{\textbf{Area}} &
      \multicolumn{1}{c|}{\textbf{T}} &
      \multicolumn{1}{c|}{\textbf{DFFs }} &
      \multicolumn{1}{c|}{\textbf{Area}} &
      \multicolumn{1}{c|}{\cellcolor[rgb]{ .949,  .949,  .949}\textbf{ADP Gain}}
      \bigstrut\\
    \hline
    Sin &
      29.4 &
      16,775 &
      118 &
      22 &
      15007 &
      113.6 &
      \cellcolor[rgb]{ .949,  .949,  .949}1.39
      \bigstrut\\
    \hline
    Max &
      28.5 &
      68,488 &
      323.3 &
      22 &
      34,833 &
      184.4 &
      \cellcolor[rgb]{ .949,  .949,  .949}2.27
      \bigstrut\\
    \hline
    Shoup &
      29.4 &
      20,707 &
      229.7 &
      22 &
      18,504 &
      222.7 &
      \cellcolor[rgb]{ .949,  .949,  .949}1.38
      \bigstrut\\
    \hline
    c2670 &
      29.4 &
      2,874 &
      16.4 &
      22 &
      2645 &
      15.8 &
      \cellcolor[rgb]{ .949,  .949,  .949}1.39
      \bigstrut\\
    \hline
    c3540 &
      29.4 &
      1,449 &
      15.8 &
      22 &
      1474 &
      16.5 &
      \cellcolor[rgb]{ .949,  .949,  .949}1.28
      \bigstrut\\
    \hline
    c5315 &
      29.4 &
      3,313 &
      24.9 &
      22 &
      2876 &
      23.4 &
      \cellcolor[rgb]{ .949,  .949,  .949}1.42
      \bigstrut\\
    \hline
    c6288 &
      29.4 &
      3,862 &
      31.4 &
      22 &
      3628 &
      31.8 &
      \cellcolor[rgb]{ .949,  .949,  .949}1.32
      \bigstrut\\
    \hline
    c7552 &
      29.4 &
      2,831 &
      23.6 &
      22 &
      3038 &
      24.9 &
      \cellcolor[rgb]{ .949,  .949,  .949}1.27
      \bigstrut\\
    \hline
    square &
      29.4 &
      44,635 &
      320.7 &
      22 &
      26,916 &
      249.4 &
      \cellcolor[rgb]{ .949,  .949,  .949}1.72
      \bigstrut\\
    \hline
    log2 &
      29.4 &
      232,214 &
      1241.9 &
      22 &
      220,360 &
      1199.5 &
      \cellcolor[rgb]{ .949,  .949,  .949}1.38
      \bigstrut\\
    \hline
    \end{tabular}%
  \label{VsFPB}%
  \end{adjustbox}
\end{table}%
%
%
%
%
%
\begin{table}[htbp]
  \centering
  \caption{Low-Area Comparison to Multi-Phase Clocking}
  \begin{adjustbox}{max width=0.9\columnwidth}
    \begin{tabular}{|l|r|r|r|r|r|r|r|}
    \hline
    \multicolumn{1}{|c|}{\multirow{3}[6]{*}{\textbf{Circuit}}} &
      \multicolumn{7}{c|}{\textbf{Method}}
      \bigstrut\\
\cline{2-8}     &
      \multicolumn{3}{c|}{\textbf{MultiPhase Clocking}} &
      \multicolumn{4}{c|}{\textbf{Delay Balancing}}
      \bigstrut\\
\cline{2-8}     &
      \multicolumn{1}{c|}{\textbf{T}} &
      \multicolumn{1}{c|}{\textbf{DFFs }} &
      \multicolumn{1}{c|}{\textbf{Area}} &
      \multicolumn{1}{c|}{\textbf{T}} &
      \multicolumn{1}{c|}{\textbf{DFFs }} &
      \multicolumn{1}{c|}{\textbf{Area}} &
      \multicolumn{1}{c|}{\cellcolor[rgb]{ .906,  .902,  .902}\textbf{ADP Gain}}
      \bigstrut\\
    \hline
    Sin &
      88.2 &
      6,296 &
      74 &
      40 &
      6562 &
      75.1 &
      \cellcolor[rgb]{ .906,  .902,  .902}2.17
      \bigstrut\\
    \hline
    Max &
      85.5 &
      28,426 &
      155.2 &
      40 &
      18,767 &
      114.6 &
      \cellcolor[rgb]{ .906,  .902,  .902}2.89
      \bigstrut\\
    \hline
    Shoup &
      88.2 &
      5,820 &
      167.2 &
      40 &
      5,230 &
      164.7 &
      \cellcolor[rgb]{ .906,  .902,  .902}2.24
      \bigstrut\\
    \hline
    c2670 &
      88.2 &
      1,228 &
      9.5 &
      40 &
      1323 &
      9.9 &
      \cellcolor[rgb]{ .906,  .902,  .902}2.12
      \bigstrut\\
    \hline
    c3540 &
      88.2 &
      462 &
      11.7 &
      40 &
      555 &
      12.1 &
      \cellcolor[rgb]{ .906,  .902,  .902}2.13
      \bigstrut\\
    \hline
    c5315 &
      88.2 &
      1,283 &
      16.4 &
      40 &
      1141 &
      15.8 &
      \cellcolor[rgb]{ .906,  .902,  .902}2.29
      \bigstrut\\
    \hline
    c6288 &
      88.2 &
      1,494 &
      21.5 &
      40 &
      1307 &
      20.7 &
      \cellcolor[rgb]{ .906,  .902,  .902}2.29
      \bigstrut\\
    \hline
    c7552 &
      88.2 &
      1,049 &
      16.1 &
      40 &
      1291 &
      17.1 &
      \cellcolor[rgb]{ .906,  .902,  .902}2.08
      \bigstrut\\
    \hline
    square &
      88.2 &
      16,235 &
      201.5 &
      40 &
      12,230 &
      184.6 &
      \cellcolor[rgb]{ .906,  .902,  .902}2.41
      \bigstrut\\
    \hline
    log2 &
      88.2 &
      79,265 &
      599.5 &
      40 &
      95,413 &
      667.3 &
      \cellcolor[rgb]{ .906,  .902,  .902}1.98
      \bigstrut\\
    \hline
    \end{tabular}%
  \label{VsMPC}%
  \end{adjustbox}
\end{table}%
\subsection{Resolving Unexpected Variations}\label{vary}
To emulate unexpected variation we took the c2670 circuit, synthesized for 22ps and performed static timing analysis on the circuit with each gate randomly receiving up to 20\% variation in their clock-to-Q delays and setup/hold time requirements.  The resulting timing violation were able to be resolved by adjusting the delay parameter of our JTL model and increasing our clock period.  Repeating this process 10x we were always able to resolve the 20\% random variations with an average JTL delay increase of 7.5\% and an average final clock period of 26.5 ps.  
\subsection{Optimizing for Variation}\label{margin}
To demonstrate our algorithms performance while optimizing for variation, we repeated the experiment from Sect. \ref{vary} at a 30ps and 40ps target clock period for a circuit that was synthesized with an expectation of $\pm$20\% variation in timing.  Since this circuit was optimized for this range of variation it did not require any post-fab tuning of delay parameters and met the desired clock period. The results in Table \ref{optvar} highlight both our ability to resolve timing variations post-fabrication and our algorithms capability to optimize for expected timing variations.
\begin{table}[h]
  \centering
  \caption{Optimizing for Variation vs Post-Fabrication Tuning}
    \begin{tabular}{|c|l|c|c|}
    \hline
    \multicolumn{4}{|c|}{\textbf{Average Performance Under 20\% Random Timing Variation}}
      \bigstrut\\
    \hline
    \multicolumn{1}{|l|}{\textbf{Circuit}} &
      \textbf{Synthesis Target} &
      \textbf{Clock Period} &
      \textbf{$\Delta$ JTL}
      \bigstrut\\
    \hline
    \multirow{4}[8]{*}{c2670} &
      30ps &
      37.4ps &
      \multicolumn{1}{r|}{10.90\%}
      \bigstrut\\
\cline{2-4}     &
      30 ps at 20\% Variation &
      30ps &
      0\%
      \bigstrut\\
\cline{2-4}     &
      40ps &
      47.5ps &
      \multicolumn{1}{r|}{9.60\%}
      \bigstrut\\
\cline{2-4}     &
      40ps at 20\% Variation &
      40ps &
      0\%
      \bigstrut\\
    \hline
    \end{tabular}%
  \label{optvar}%
\end{table}%
\section{Conclusions}
SFQ circuit and fabrication technology continues to improve, driving the necessity for strong EDA algorithms that can utilize the full potential of superconducting devices.  We utilize the clocked nature of logic gates and DFFs to control the delays along datapaths, allowing fine control of data sequencing to achieve high performance with lower area.  We also proposed methods that provide the ability to tune these delays post-fabrication leading to robustness to the timing variations that often plague SFQ designs. Our designs demonstrated an average ADP improvement of 2.07x over state of the art area reduction designs and a 1.48X ADP improvement over high frequency FPB designs.  Our work enables designers to consider a range of performance-area trade-offs while allowing the circuit to be modified after fabrication to accommodate unexpected timing variation.  Further work can be done to explore optimized placement and routing of delay balanced designs given knowledge of which connections are on critical paths.  
\bibliographystyle{IEEEtran}
\bibliography{bibliography}



\end{document}